\documentclass[preprint,review,12pt]{elsarticle}

\usepackage{amssymb}
\usepackage{amsmath}
\usepackage{subfigure}
\usepackage{array}
\usepackage{multirow}
\usepackage{threeparttable}
\biboptions{numbers,sort&compress}
\usepackage{float}
\usepackage{subcaption}

\journal{Signal Processing}

\begin{document}

\begin{frontmatter}



\title{Graph Linear Canonical Transform Based on CM-CC-CM Decomposition}


\author[1,2,3]{Na Li}
\author[1,2,3]{Zhichao Zhang \corref{cor1}}
\ead{zzc910731@163.com}
\author[4]{Jie Han}
\author[1,2,3]{\\ Yunjie Chen}
\author[1,2,3]{Chunzheng Cao}


\address[1]{School of Mathematics and Statistics, Nanjing University of Information Science and Technology, Nanjing 210044, China}
\address[2]{Center for Applied Mathematics of Jiangsu Province, Nanjing University of Information Science and Technology, Nanjing 210044, China}
\address[3]{Jiangsu International Joint Laboratory on System Modeling and Data Analysis, Nanjing University of Information Science and Technology, Nanjing 210044,
China}
\address[4]{School of Remote Sensing and Geomatics Engineering, Nanjing University of Information Science and Technology, Nanjing 210044, China}

\cortext[cor1]{Corresponding author}
\tnotetext[]{This work was supported by the National Natural Science Foundation of China under Grant No 61901223 and the Jiangsu Planned Projects for Postdoctoral Research Funds under Grant No 2021K205B.}


\begin{abstract}
The graph linear canonical transform (GLCT) is presented as an extension of the graph Fourier transform (GFT) and the graph fractional Fourier transform (GFrFT), offering more flexibility as an effective tool for graph signal processing. In this paper, we introduce a GLCT based on chirp multiplication-chirp convolution-chirp multiplication decomposition (CM-CC-CM-GLCT), which irrelevant to sampling periods and without oversampling operation. Various properties and special cases of the CM-CC-CM-GLCT are derived and discussed. In terms of computational complexity, additivity, and reversibility, we compare the CM-CC-CM-GLCT and the GLCT based on the central discrete dilated Hermite function (CDDHFs-GLCT). Theoretical analysis demonstrates that the computational complexity of the CM-CC-CM-GLCT is significantly reduced. Simulation results indicate that the CM-CC-CM-GLCT achieves similar additivity to the CDDHFs-GLCT. Notably, the CM-CC-CM-GLCT exhibits better reversibility.

\end{abstract}




\end{frontmatter}


\section{Introduction}
\label{}
In 1994, Almeida \cite{Almeida1994TheFF} elucidated the rotational properties of the fractional Fourier transform (FrFT). The incorporation of the rotation angle parameter allows signal representation in the fractional domain situated between the time and frequency domains. The optimal representation of signals in fractional domains is achieved by analyzing the signals across all fractional domains, encompassing both the time and frequency domains \cite{Candan1999TheDF,Shi2020NovelSF,Farah2020ANC}. Linear canonical transform (LCT) expands rotation to affine transformation, thereby extending the Fourier transform (FT) and FrFT as a more expansive parameterized linear integral transform. In the 1970s, Professors Collin \cite{Collins1970LensSystemDI} and Moshinsky \cite{Moshinsky1971LinearCT} proposed the concept of LCT, while complex parameter LCT was previously discussed by Bargmann. Compared with FT and FrFT, LCT introduces three independent parameters, enabling non-band-limited signals in traditional FT and FrFT domains to maybe become band-limited signals in the LCT domain under specific parameter settings. Unlike the rotational properties of FrFT in the time-frequency plane, LCT exhibits affine transformation characteristics that encompass compression and stretching operations. Therefore, the LCT offers increased flexibility and freedom, possesses greater potential in signal processing, and enables improved signal representation \cite{Shi2018LinearCM,Wei2021SparseDL}.

To facilitate the effective application of the LCT in signal processing and optics, it is essential to investigate the discretization methods and fast algorithms specific for LCT. Developing an optimal discrete algorithm while preserving characteristics of continuous transformations and minimizing computational complexity is essential for achieving accurate numerical computations of LCT. In this case, the fast discretization algorithm for LCT has consistently been a focal point in LCT research. Based on the current research situation, the fundamental ideas of LCT discretization are mainly divided into three types. The first is the direct discretization method \cite{Zhao2008SamplingRC, Hennelly2005FastNA, Zhao2013UnitaryDL}. By drawing from the method of the classical discrete-time Fourier transform, the definition and characteristics of the discrete linear canonical transform (DLCT) are given through an emulation of imitating the process from continuous Fourier transform to discrete Fourier transform (DFT). Zhao et al. \cite{Zhao2008SamplingRC} introduced the definition of DLCT through direct discretization and applied it to sampling rate conversion research based on DLCT. The second method involves the LCT decomposition utilizing the fast Fourier transform (FFT) algorithm. Based on the direct discretization method to define the DLCT, an $N$-point DLCT can be decomposed into multiple shorter sequences of DLCT.  Healy et al. \cite{Healy2006NewFA,Healy2010FastLC} proposed a fast unitary basis decomposition algorithm leveraging FFT principles and the properties of discrete transformation matrices. At the same time, a fast algorithm for parameter matrix decomposition is presented in accordance with the definition and additivity of LCT, effectively approximating continuous LCT. The third method pertains to the decomposition of the parameter matrix \cite{Ko2008DigitalCO,Pei2011DiscreteLC,Serbes2011TheDF,Campos2009AFA}. Drawing from the decomposition property of the LCT parameter matrix, it is possible to break down the parameter matrix into the product of several special matrices. Consequently, the LCT is resolved into the product of familiar transformations, such as FT, FrFT, scaling transformation, chirp convolution (CC), and chirp multiplication (CM). By employing the existing algorithms for these transformations, rapid computation of the LCT becomes feasible, facilitating a close approximation of the continuous LCT. The key of this method lies in determining the sampling rate and sampling points. Ensuring adequate sampling in the decomposition process is essential to effectively reconstruct continuous transformations from discrete points. Koc et al. \cite{Ko2008DigitalCO} introduced the decomposition of most parameter matrices, leading to the proposal of various discretization and fast calculation methods based on different signal types and decomposition techniques. Pei et al. \cite{Pei2011DiscreteLC} presented a DLCT based on Iwasawa decomposition without oversampling. The Hermite function serves as the fundamental function of FrFT, enabling the realization of the discrete fractional Fourier transform (DFrFT) through the orthogonal Hermite function basis. Iwasawa decomposition involves scaling transformation, thereby defining the DLCT in terms of the scalable discrete Hermite functions derived from the central discrete dilated Hermite function (CDDHFs).

In traditional signal processing, signals are typically represented in regular format. However, with the rapid development of information technology, signals encountered in practical problems frequently involve complex high-dimensional data. These signals has an irregular topological structure distinct from the conventional time-space domain, so requires new processing methods. Therefore, the emergence of graph signal processing technology occurred \cite{Shuman2012TheEF, Sandryhaila2014BigDA, Li2022DistributedFL, Yang2021ARM}. Graph signal processing extends conventional discrete signal processing to graph signals with topological structures, offering an effective approach for handling data with intricate structures. This technology finds widespread application in radar surveillance, biomedicine, image processing, and other domains. In order to study graph signals, it is necessary to extend the concepts from traditional signal processing to the graph signal domain. This includes fundamental concepts such as graph signal transformation, convolution, filtering, shifting, and modulation, which collectively establish a complete theoretical system for graph signal processing. Graph signal processing is relatively new and still lacks several vital concepts and theories, one of which is the graph linear canonical transform (GLCT). The GLCT, like the LCT, represents a versatile multi-parameter linear integral transformation with enhanced flexibility, making it a significant tool in signal processing. Despite its importance, the proportion of theoretical works defining GLCT remains limited. In a recent study, Zhang and Li \cite{Zhang2022DiscreteLC} introduced a definition of CDDHFs-GLCT, achieved through a combination of graph CM \cite{Pei2002EigenfunctionsOL, Satyan2010ChirpMB}, graph scaling transformation, and graph fractional Fourier transform (GFrFT) \cite{Yan2021MultidimensionalGF, Zhang2023TheFF, Wei2023SamplingOG}. While this discrete algorithm exhibits favorable properties, it is associated with high computational complexity, thereby posing challenges for practical application. Thus, the practical application of existing methods is limited. To address this limitation, this paper introduces CM-CC-CM-GLCT that irrelevant to sampling periods and without oversampling operation. The main contributions of this paper are summarized as follows:  \\
$\bullet$ The paper proposes the CM-CC-CM-GLCT based on the CM-CC-CM decomposition without scaling transformation, which avoids the possibility of changing the sampling period or causing the interpolation error. \\
$\bullet$ In terms of computational complexity, CM-CC-CM-GLCT exhibits lower computational complexity than CDDHFs-GLCT, which also irrelevant to sampling periods. It has higher efficiency and performance in practical applications. \\
$\bullet$ In terms of additivity and reversibility, CM-CC-CM-GLCT can achieve similar additivity compared with CDDHFs-GLCT. Most importantly, CM-CC-CM-GLCT has better reversibility, which offers an auxiliary thinking in some applications such as coding and decoding.

The structure of this paper is as follows. In Section 2, briefly proposes the graph signal theory on which this study is based. In Section 3, gives a review of the CDDHFs-GLCT. In Section 4, we define CM-CC-CM-GLCT, and discuss some properties and special cases. In Section 5, CM-CC-CM-GLCT is compared with CDDHFs-GLCT in terms of computational complexity, additivity and reversibility. Finally, the paper is summarized in Section 6.


\section{Preliminaries}
\label{}

Consider a graph $G\left ( U,E \right )$, where $U$ represents the set of $N$ nodes in the graph, and $E$ is the set of edges. In this paper, the fundamental structure of a graph $G$ be captured by its adjacency matrix $\mathbf{A}$, where each data element corresponds to one edge. Specifically, for any two different nodes $u_{i},u_{j}\in U$, $\mathbf{A}\left ( i,j \right )=\omega _{ij} $, with $\omega _{ij}  $ represents the connection weight between node $u_{i}$ and $u_{j}$. Given a graph $G$, the data set consists of graph signals, defined as a mapping, $\mathbf{x}:u_{n}\longrightarrow x_{n} $. Then the graph signal can be expressed as a vector $\mathbf{x}=\left [ x_{0},x_{1}\dots ,x_{N-1}   \right ]^{\rm{T}}$, which is not only a list, but also a graph.

There are two main construction frameworks for graph signal processing, which are based on spectral method and algebraic method respectively. The first framework is based on the construction of Laplace matrix, and intuitively analyzes the graph structure based on spectrogram theory. Specifically, the graph Fourier transform (GFT) is defined by the eigenfunction of Laplacian matrix, and the corresponding spectrum are represented by eigenvalues. Since the Laplacian matrix must be symmetric and positive semidefinite, this method is exclusively applicable to undirected graphs. The second framework is derived from the discrete signal processing on graphs, which analyzes the characteristics of graph structure signals according to algebraic methods. Specifically, GFT is defined by using the eigenfunction of adjacency matrix, and the corresponding spectrum are represented by eigenvalues. Since the adjacency matrix may not be symmetric, this method is applicable to any graph. This paper defines GFT based on the adjacency matrix.

For the convenience of discussion, it is assumed that the adjacency matrix $\mathbf{A}$ is symmetric. Its eigendecomposition is $\mathbf{A=V\Lambda _{A}V^{-1}}$, where $\mathbf{\Lambda _{A}}$ is a diagonal matrix composed of eigenvalues $\left \{ \lambda _{0} , \lambda _{1},\dots,\lambda _{N-1} \right \}$ , $\mathbf{V}$ is a matrix composed of eigenvectors $\left \{ \mathbf{v} _{0} , \mathbf{v} _{1},\dots,\mathbf{v} _{N-1} \right \}$, and $\mathbf{V}^{-1} $ is the inverse matrix of $\mathbf{V}$. Eigenvalues and eigenvectors contain the information of graph structure signals. Sort the eigenvalues from small to large, $\lambda _{0}\le\lambda _{1}\le \dots \le \lambda _{N-1} $. Let $\mathbf{\lambda} =\left \{ \lambda _{0},\lambda _{1}, \dots , \lambda _{N-1}  \right \}$, then $\mathbf{\lambda}$ is called the spectrum of $G$ and $\lambda_{i}$ is called the frequency. The larger the eigenvalues, the more high-frequency information contained in the corresponding eigenvectors, and the more violent the fluctuations. Conversely, the eigenvector with smaller eigenvalues corresponds to the low-frequency eigenvector, which describes the smooth characteristics in the graph signal and has reduced fluctuation.

Definition 1. Given a graph $G$. $\mathbf{A}$ is an adjacency matrix, and the eigen-decomposition is $\mathbf{A=V\Lambda _{A}V^{-1}}$. The graph Fourier transform of signal $\mathbf{x}$ is defined as
\begin{equation}
\mathbf{\hat{x}}=\mathbf{Fx}=\mathbf{V}^{-1}\mathbf{x},  
\end{equation}
where $\mathbf{F}=\mathbf{V}^{-1} $ is the graph Fourier transform matrix. $\hat{\mathbf{x}}=\left [ \hat{x }_{0},\hat{x }_{1},\dots ,\hat{x }_{N-1}\right ]^{\rm{T}}$.

Definition 2. Given a graph $G$. $\mathbf{A}$ is an adjacency matrix, and the eigendecomposition is $\mathbf{A=V\Lambda _{A}V^{-1}}$. The graph fractional Fourier transform of signal $\mathbf{s}$ is defined as
\begin{equation}
	\mathbf{\hat{s}}=\mathbf{F^{\alpha }s }=\mathbf{V}^{-\alpha }\mathbf{s},  
\end{equation}
where parameter $\alpha $ is the order of graph fractional Fourier transform.

If $\mathbf{F}$ is ortho-diagonalized, then $\mathbf{F}=\mathbf{V}^{-1}=\mathbf{Q\Lambda Q}^{\rm{T}}$. The graph fractional Fourier transform of signal $\mathbf{s}$ is defined as
\begin{equation}
	\mathbf{\hat{s}}=\mathbf{F^{\alpha }s }=\mathbf{Q\Lambda^{\alpha }  Q}^{\rm{T}}\mathbf{s},
\end{equation}
where $\alpha \in \left [ 0,1 \right ]$.

\section{ Review of CDDHFs-GLCT }
In this section, we will briefly review CDDHFs-GLCT.

Definition 3. Let the parameter matrix $\mathbf{M}=(a,b,c,d),a,b,c,d\in {\mathbb{R}},ad-bc=1$. The linear canonical transform of signal $x\left ( t \right ) $ with the parameter matrix $\mathbf{M}$ is defined as
\begin{equation}
	O_{\rm{LCT}}^{\left ( a,b,c,d \right ) }\left \{ x\left ( t \right)\right \} =
\left\{\begin{array}{lcl}
	\sqrt{\frac{1}{ib} }\int_{-\infty }^{\infty }e^{i2\pi \left ( \frac{d}{2b}u^{2}-\frac{1}{b}ut+\frac{a}{2b}t^{2}\right )}x\left ( t \right )\rm{dt}    
	 &   &{b\ne 0}\\
	\sqrt{d}e^{i\pi cdu^{2} }x\left ( du \right )   
	 &   &{b=0}
\end{array}
,\right  
.\end{equation}
where $O_{\rm{}}^{\left ( a,b,c,d \right ) }$ represents LCT operator.

Graph signal is a discrete signal with topological structure. GLCT serves as an extension of the DLCT in graph signal processing, similar to GFT extends DFT. Therefore, we need to give the definition of DLCT first, which is the process of discretization of LCT. There exist three methods for discretization and fast algorithm development for LCT: direct discretization based on classical discrete-time Fourier transform, linear canonical transformation decomposition based on FFT algorithm principle and parameter matrix decomposition. The CDDHFs-GLCT adopts CDDHFs decomposition in \cite{Zhang2022DiscreteLC}.

The DLCT is based on CDDHFs decomposition:
\begin{equation}
\mathbf{M} = \begin{pmatrix}
	a&b \\
	c&d
\end{pmatrix} = \begin{pmatrix}
	1&0 \\
	\xi &1
\end{pmatrix}
\begin{pmatrix}
	\sigma &0 \\
	0&\sigma ^{-1} 
\end{pmatrix}
\begin{pmatrix}
	\cos\alpha  &\sin\alpha   \\
	-\sin \alpha &\cos\alpha  
\end{pmatrix}.
\end{equation}

The first matrix corresponds to CM with chirp rate $\xi $. The second matrix corresponds to a scaling transformation with scaling factor $\delta$. The third matrix corresponds to DFrFT with transformation order $\alpha$. The relationship between parameter $\left ( \xi ,\delta ,\alpha  \right )$ and parameters $(a,b,c,d)$ is
\begin{equation}
	\xi =\frac{ac+bd}{a^{2}+b^{2}},\delta =\sqrt{a^{2}+b^{2}},\alpha =\cos ^{-1}\left ( \frac{a}{\delta }  \right )= \sin  ^{-1}\left ( \frac{b}{\delta }  \right ).
\end{equation}
This decomposition is a special case of Iwasawa decomposition \cite{delaCruz2021OnTI} of symplectic matrix. Under this decomposition, DLCT can be realized by the combination of CM, scaling transform and DFrFT.

For discrete signal $\left \{ x\left [ n \right ]  \right \}$, DFrFT is obtained by eigendecomposition:
\begin{equation}
	O_{\rm{CDDHFs-DLCT}}^{\left ( \cos \alpha ,\sin \alpha ,-\sin \alpha,\cos \alpha   \right ) }\left ( x\left [ n \right ]  \right )=\mathbf{ED}_{\alpha }\mathbf{E}^{\rm{T}}x\left [ n \right ], 
\end{equation}
\begin{equation}
	\mathbf{D}_{\alpha } \left [ m,n \right ]=\left\{
	\begin{array}{lcl}
		{\rm e}^{-{\rm i}\left ( m+\frac{1}{2}\alpha   \right ) }  &   &m=n \quad and \quad 0\le m,n\le N-1\\
		0&   &\rm{otherwise}
	\end{array}
	,\right
.\end{equation}
where $\mathbf{E}$ is an $N\times N$ orthonormalsquare matrix with $p$th column being the $p$th order CDDHFs with $\sigma =1  $, and $\mathbf{D}_{\alpha }$ is a diagonal matrix.

For discrete signal $\left \{ x\left [ n \right ]  \right \}$, the discrete scaling transform process represents
\begin{equation}
	O_{\rm{CDDHFs-DLCT}}^{(\delta ,0,0,\delta ^{-1} )}\left ( x\left [ n \right ]  \right )=\mathbf{E}_{\delta }\mathbf{E}^{\rm{T}}x\left [ n \right ],
\end{equation}
where $\mathbf{E}_{\delta }$ is an $N\times N$ orthonormalsquare matrix with $p$th column being the $p$th order CDDHFs, $\delta$ is the scaling transform factor.

For discrete signal $\left \{ x\left [ n \right ]  \right \}$, the discrete CM represents
\begin{equation}
	O_{\rm{CDDHFs-DLCT}}^{(1,0,\xi ,1 )}\left ( x\left [ n \right ]  \right )=\mathbf{D}_{\xi } x\left [ n \right ],
\end{equation}
where $\mathbf{D}_{\xi }$ is chirp diagonal matrix.

Based on the above three steps, for discrete signal $\left \{ x\left [ n \right ]  \right \}$, DLCT based on the decomposition formula (5) represents
\begin{equation}
	\begin{array}{lcl}
		O_{\rm{CDDHFs-DLCT}}^{(a,b,c,d )}\left ( x\left [ n \right ] \right )& = & \left \{ \left ( \mathbf{D}_{\xi }  \right )\left (\mathbf{ E}_{\delta } \mathbf{E}^{\rm{T}}  \right ) \left ( \mathbf{ED}_{\alpha }\mathbf{E}^{\rm{T}} \right )   \right \}x\left [ n \right ]  \\
		& = & \left \{ \mathbf{D}_{\xi }\mathbf{E}_{\delta }\mathbf{D}_{\alpha }\mathbf{E}^{\rm{T}}     \right \}x\left [ n \right ] 
	\end{array}.
\end{equation}

On the basis that CDDHFs is the eigenfunctions of DFrFT and scaling transform, the CDDHFs-DLCT is defined in \cite{Zhang2022DiscreteLC}. According to the decomposition formula (5), CDDHFs-GLCT can also be realized through three stages: graph CM, graph scaling transform and GFRFT.

For graph signal $\mathbf{x}$, GFrFT is obtained by eigendecomposition:
\begin{equation}
	O_{\rm{CDDHFs-GLCT}}^{\left ( \cos \alpha ,\sin \alpha ,-\sin \alpha,\cos \alpha   \right ) }\left ( \mathbf{x} \right )  =\mathbf{F}^{\alpha}\mathbf{x}=\mathbf{Q}\mathbf{\Lambda }^{\alpha }\mathbf{Q}^{\rm{T}}\mathbf{x}, 
\end{equation}
where $\alpha$ is the rotation angle.

In the graph scaling operator stage, the adjacency matrix $\mathbf{A}$ is directly scaled to obtain the graph shift operator $\mathbf{S}=\frac{1}{\delta }\mathbf{A}$. The graph spectrum decomposition is $\mathbf{S}=\mathbf{Q}_{\delta }\mathbf{\Lambda _{S}Q}_{\delta }^{\rm{T}}$. Where $\delta$ is the scaling transform factor, $\mathbf{Q}_{\delta }$ and $\mathbf{\Lambda _{S}}$ represent the eigenvectors of matrix $\mathbf{S}$ and their corresponding eigenvalues, respectively. For graph signal $\mathbf{x}$, the graph scaling transform process represents
\begin{equation}
		O_{\rm{CDDHFs-GLCT}}^{(\delta ,0,0,\delta ^{-1} )} \left ( \mathbf{x} \right ) =\mathbf{Q}_{\delta }\mathbf{\Lambda _{S}Q}_{\delta }^{\rm{T}}\mathbf{x}.
\end{equation}

In graph signal processing, there are few theories to define the chirp signals on graphs. Therefore, the graph CM matrix is defined in terms of the chrip Fourier transform on graphs in \cite{Zhang2022DiscreteLC}. For graph signal $\mathbf{x}$, the graph CM represents
\begin{equation}
		O_{\rm{CDDHFs-GLCT}}^{(1,0,\xi ,1 )}\left ( \mathbf{x} \right )=\mathbf{F}^{\xi }\mathbf{x}=\mathbf{Q\Lambda} ^{\xi }\mathbf{Q}^{\rm{T}}\mathbf{x},
\end{equation}
where $\xi $ is the graph CM parameter and $\mathbf{\Lambda} ^{\xi }$ is the eigenvalue of the matrix $\mathbf{A}$ after CM.

Based on the above three steps, for graph signal $\mathbf{ x}$, CDDHFs-GLCT represents
\begin{equation}
	\begin{array}{lcl}
		O_{\rm{CDDHFs-GLCT}}^{(a,b,c,d )}\left ( \mathbf{x} \right )& = & \left \{ \Lambda ^{\xi }(\mathbf{Q}_{\delta }\mathbf{\Lambda _{S}Q}_{\delta }^{\rm{T}}) (\mathbf{Q}\mathbf{\Lambda }^{\alpha }\mathbf{Q}^{\rm{T}}) \right \}\mathbf{x}  \\
		& = & \left \{ \Lambda ^{\xi }\mathbf{Q}_{\delta }\mathbf{\Lambda }^{\alpha }\mathbf{Q}^{\rm{T}} \right \}\mathbf{x} 
	\end{array}.
\end{equation}

\section{ CM-CC-CM-GLCT }

While CDDHFs decomposition offers a lower oversampling rate compared to the direct discrete method, the presence of scaling transformations poses the risk of altering the sampling period or introducing interpolation errors. To overcome this problem, the paper adopts for the CM-CC-CM decomposition method, which circumvents scaling transformations. The CM-CC-CM-GLCT not only irrelevant to sampling periods and without oversampling operation but also significantly reduces computational complexity.

\subsection{Formulation of the CM-CC-CM-GLCT for $b\ne 0$}
We first introduce CM-CC-CM decomposition:
\begin{equation}
	\mathbf{M} = \begin{pmatrix}
		a&b \\
		c&d
	\end{pmatrix} = \begin{pmatrix}
		1&0 \\
		\frac{d-1}{b} &1
	\end{pmatrix}
	\begin{pmatrix}
		1 &b \\
		0&1 
	\end{pmatrix}
	\begin{pmatrix}
		1  &0   \\
		\frac{a-1}{b} &1  
	\end{pmatrix}.
\end{equation}
The first matrix corresponds to CM with chirp rate $\frac{d-1}{b}$. The second matrix corresponds to CC with parameter $b$. The third matrix corresponds to CM with chirp rate $\frac{a-1}{b}$. The CC can be further decomposed, that is 
\begin{equation}
	\mathbf{M} = \begin{pmatrix}
		1&b \\
		0&1
	\end{pmatrix} = \begin{pmatrix}
		0&-1 \\
		1&0
	\end{pmatrix}
	\begin{pmatrix}
		1 &0 \\
		-b&1 
	\end{pmatrix}
	\begin{pmatrix}
		0 &1   \\
		-1 &0  
	\end{pmatrix}.
\end{equation}
Then
\begin{equation}
\mathbf{M}  = \begin{pmatrix}
		1&0 \\
		{\xi _{1} } &1
	\end{pmatrix}
	\begin{pmatrix}
		0&-1 \\
		1&0
	\end{pmatrix}
	\begin{pmatrix}
		1 &0 \\
		{\xi _{2} }&1 
	\end{pmatrix}
	\begin{pmatrix}
		0 &1   \\
		-1 &0  
	\end{pmatrix}
	\begin{pmatrix}
		1  &0   \\
		{\xi _{3} } &1  
	\end{pmatrix}.
\end{equation}
The relationship between parameter $\left (\xi _{1} , \xi _{2} , \xi _{3 } \right )$ and parameters $(a,b,c,d)$ is
\begin{equation}
	\xi _{1}=\frac{d-1}{b},\xi _{2}=-b,\xi _{3}=\frac{a-1}{b}.
\end{equation}
We obtain CM-CC-CM-DLCT composed of DFT, IDFT and discrete CM \cite{Pei2016FastDL}:
\begin{equation}
	O_{\rm{CM-CC-CM-DLCT} }^{\left ( a,b,c,d \right )}\left ( x\left [ n \right ] \right )  =\left \{ \mathbf{D}_{\xi _{1} }\mathbf{F}^{-1}\mathbf{D}_{\xi _{2} }\mathbf{FD}_{\xi _{3} }     \right \}x\left [ n \right ],    
\end{equation}
where $\mathbf{D}_{\xi _{1} }, \mathbf{D}_{\xi _{2} }, \mathbf{D}_{\xi _{3} }$ are diagonal chirp matrices, $\mathbf{F}$ represents DFT matrix, and $\mathbf{F}^{-1}$ represents IDFT matrix. Since DFT and IDFT can be efficiently computed by FFT, their computational complexity is obviously lower than that of CDDHFs decomposition.

GLCT is an extension of DLCT in graph signal processing. The CM-CC-CM-GLCT can also be realized by the combination of graph CM, GFT and IGFT. According to the definition of graph CM above, the CM-CC-CM-GLCT represents
\begin{equation}
	O_{\rm{CM-CC-CM-GLCT} }^{\left ( a,b,c,d \right )}\left (\mathbf{ x} \right ) = \left \{ \mathbf{\Lambda} ^{\xi _{1} }\mathbf{F}^{-1}\mathbf{\Lambda} ^{\xi _{2} }\mathbf{F\Lambda} ^{\xi _{3} }   \right \}\mathbf{x}.
\end{equation}

\subsection{Formulation of the CM-CC-CM-GLCT for $b=0$}
By observing the parameters, we can find that the definition in formula (21) is invalid when $b=0$. In the case of $b=0$, since the parameter satisfies $ad-bc=1$, there are $a\ne 0$ and $d\ne 0$. Then the following two kinds of decomposition can be considered \cite{Pei2016FastDL}:
\begin{equation}
	\mathbf{M}  = \begin{pmatrix}
		a&0 \\
		c &d
	\end{pmatrix}=
	\begin{pmatrix}
		0&1 \\
		-1&0
	\end{pmatrix}
	\begin{pmatrix}
		1 &0 \\
		\eta _{1}&1 
	\end{pmatrix}
	\begin{pmatrix}
		0&-1 \\
		1&0
	\end{pmatrix}
	\begin{pmatrix}
		1 &0 \\
		\eta _{2}&1 
	\end{pmatrix}
	\begin{pmatrix}
		0 &1   \\
		-1 &0  
	\end{pmatrix}
	\begin{pmatrix}
		1  &0   \\
		\eta _{3} &1  
	\end{pmatrix},
\end{equation}
and
\begin{equation}
	\mathbf{M}  = \begin{pmatrix}
		a&0 \\
		c &d
	\end{pmatrix}=
	\begin{pmatrix}
		1&0 \\
		\mu _{1}&1
	\end{pmatrix}
	\begin{pmatrix}
		0&-1 \\
		1&0
	\end{pmatrix}
	\begin{pmatrix}
		1 &0 \\
		\mu_{2}&1 
	\end{pmatrix}
	\begin{pmatrix}
		0 &1   \\
		-1 &0  
	\end{pmatrix}
		\begin{pmatrix}
		1&0 \\
		\mu_{3 }&1
	\end{pmatrix}
	\begin{pmatrix}
		0  &-1  \\
		1 &0  
	\end{pmatrix}.
\end{equation}
The relationship between parameter $\left (\eta _{1} , \eta _{2} , \eta _{3 }  \right )$ and parameters $(a,b,c,d)$ is
\begin{equation}
	\eta _{1}=\frac{1}{d},\eta _{2}=d,\eta _{3}=\frac{c+1}{d}.
\end{equation}
The relationship between parameter $\left (\mu _{1} , \mu _{2} , \mu _{3 }  \right )$ and parameters $(a,b,c,d)$ is
\begin{equation}
	\mu _{1}=\frac{c-1}{a},\mu _{2}=-a,\mu _{3}=-\frac{1}{a}.
\end{equation}
The matrix forms of DLCT based on (22) and (23) are given by the following formulas respectively:
\begin{equation}
	O_{\rm{CM-CC-CM-DLCT} }^{\left ( a,,0,c,d \right )}\left ( x\left [ n \right ]  \right )  =\left \{ \sqrt{-\rm{j}}\mathbf{FD}_{\eta _{1} }\mathbf{F}^{-1}\mathbf{D}_{\eta _{2} }\mathbf{FD}_{\eta _{3} } \right \}x\left [ n \right ], 
\end{equation}
\begin{equation}
	O_{\rm{CM-CC-CM-DLCT} }^{\left ( a,0,c,d \right )}\left ( x\left [ n \right ]  \right ) =\left \{ \sqrt{-\rm{j}}\mathbf{D}_{\mu _{1} }\mathbf{F}^{-1}\mathbf{D}_{\mu _{2} }\mathbf{FD}_{\mu _{3} }\mathbf{F}^{-1}  \right \}x\left [ n \right ],
\end{equation}
where $\sqrt{\rm{j}}$ is the phase difference between FT and LCT, and $\sqrt{-\rm{j}}$ is the phase difference between IFT and LCT.
When $b=0$, the CM-CC-CM-GLCT is given by the following formula:
\begin{equation}
	O_{\rm{CM-CC-CM-GLCT} }^{\left ( a,0,c,d \right )}\left ( \mathbf{x} \right )  =\left \{ \sqrt{-\rm{j}}\mathbf{F\Lambda} ^{\eta  _{1} }\mathbf{F}^{-1}\mathbf{\Lambda} ^{\eta _{2} }\mathbf{F\Lambda} ^{\eta _{3} } \right \} \mathbf{x},                 
\end{equation}
\begin{equation}
O_{\rm{CM-CC-CM-GLCT} }^{\left ( a,,0,c,d \right )}\left ( \mathbf{x} \right )  =\left \{ \sqrt{\rm{j}}\mathbf{\Lambda} ^{\mu _{1} }\mathbf{F}^{-1}\mathbf{\Lambda} ^{\mu _{2} }\mathbf{F\Lambda} ^{\mu _{3} }\mathbf{F}^{-1} \right \} \mathbf{x} .
\end{equation}

\subsection{Properties}
In this subsection, we introduce some properties of CM-CC-CM-GLCT, including zero rotation, additivity, reversibility and unitary.

$\left ( 1 \right )$Zero rotation
\begin{equation}
	O_{\rm{CM-CC-CM-GLCT} }^{\left ( 1,0,0,1 \right ) }=\mathbf{\Lambda} ^{0}\mathbf{F}^{-1}\mathbf{\Lambda} ^{0}\mathbf{F\Lambda} ^{0} =\mathbf{F^{-1}}\mathbf{F}=\mathbf{I},
\end{equation}
where $\begin{pmatrix}
	a&b \\
	c &d
\end{pmatrix}=
\begin{pmatrix}
	1&0 \\
	0&1
\end{pmatrix}$, and calculated $\xi _{1}=0,\xi _{2}=0,\xi _{3}=0$.

$\left ( 2 \right )$Additivity

The additivity property of the GLCT is expressed as matrix multiplication, and the additivity property of the CM-CC-CM-GLCT represents
\begin{equation}
	\mathbf{\Lambda} ^{\xi _{1}^{3} }\mathbf{F}^{-1}\mathbf{\Lambda} ^{\xi _{2}^{3} }\mathbf{F\Lambda} ^{\xi _{3}^{3} }\left ( \mathbf{x} \right )=\mathbf{\Lambda} ^{\xi _{1}^{2} }\mathbf{F}^{-1}\mathbf{\Lambda} ^{\xi _{2}^{2} }\mathbf{F\Lambda} ^{\xi _{3}^{2} }\left \{ \mathbf{\Lambda }^{\xi _{1}^{1} }\mathbf{F}^{-1}\mathbf{\Lambda} ^{\xi _{2}^{1} }\mathbf{F\Lambda} ^{\xi _{3}^{1} }\left (\mathbf{ x} \right )  \right \} ,
\end{equation}
where $\begin{pmatrix}
	a_{3}&b_{3} \\
	c_{3} &d_{3}
\end{pmatrix}=
\begin{pmatrix}
	a_{2}&b_{2} \\
	c_{2} &d_{2}
\end{pmatrix}
\begin{pmatrix}
a_{1}&b_{1} \\
c_{1} &d_{1}
\end{pmatrix}$. It implies that the cascade of several GLCTs with parameter matrices can be replaced by only one GLCT using parameter matrix.

$\left ( 3 \right )$ Reversibility

The reversibility property is a special case of the additivity property. The reversibility allows the IGLCT with parameter matrix to be realized by the forward GLCT with parameter matrix. The reversibility property of the CM-CC-CM-GLCT is given by
\begin{equation}
	{\rm inv}\left ( O_{\rm{CM-CC-CM-GLCT} }^{\left ( a,b,c,d \right ) }  \right )= \left ( O_{\rm{CM-CC-CM-GLCT} }^{\left ( a,b,c,d \right ) }  \right )^{-1}=\mathbf{\Lambda}^{-\xi _{3} }\mathbf{F}^{-1}\mathbf{\Lambda} ^{-\xi _{2} }\mathbf{F\Lambda} ^{-\xi _{1} },
\end{equation}
where ${\rm inv}\left ( \bullet  \right )$ represents the inverse matrix. Another approach to represent IGLCT is recalculating the parameters of another GLCT using the inverse matrix of the original GLCT. The reversibility property of the CM-CC-CM-GLCT is given by
\begin{equation}
	{\rm inv}\left ( O_{\rm{CM-CC-CM-GLCT} }^{\left ( a,b,c,d \right ) }  \right )= O_{\rm{CM-CC-CM-GLCT} }^{\left ( d,-b,-c,a \right ) }=\mathbf{\Lambda} ^{\hat{\xi }_{1}  }\mathbf{F}^{-1}\mathbf{\Lambda} ^{\hat{\xi }_{2}  }\mathbf{F\Lambda} ^{\hat{\xi}_{3}  }.
\end{equation}

$\left ( 4 \right )$ Unitary

The adjacency matrix $\mathbf{A}$ of a real weighted undirected graph is a real symmetric matrix, so $\mathbf{A}$ can be diagonally orthogonal:
\begin{equation}
	\mathbf{A}=\mathbf{V\Lambda _{A}V}^{-1}=\mathbf{V\Lambda _{A}V}^{\rm{T}},
\end{equation}
where $\mathbf{F=V}^{\rm{T}}$ is an orthogonal matrix. Since the eigenvalue modulus of orthogonal matrix is 1, then
\begin{equation}
		\begin{array}{lcl}
			O_{\rm{CM-CC-CM-GLCT} }^{\left ( a,b,c,d \right ) }\left ( O_{\rm{CM-CC-CM-GLCT} }^{\left ( a,b,c,d \right ) } \right )^{\rm{H}}   \\ 
           = \left ( \mathbf{\Lambda} ^{\xi _{1} }\mathbf{F}^{-1}\mathbf{\Lambda} ^{\xi _{2} }\mathbf{F\Lambda} ^{\xi _{3} }     \right ) \left ( \mathbf{\Lambda} ^{\xi _{1} }\mathbf{F}^{-1}\mathbf{\Lambda} ^{\xi _{2} }\mathbf{F\Lambda} ^{\xi _{3} }     \right )^{\rm{H}}  \\
		 = \mathbf{\Lambda }^{\xi _{1} }\mathbf{F}^{-1}\mathbf{\Lambda}^{\xi _{2} }\mathbf{F\Lambda}^{\xi _{3} } \mathbf{\Lambda} ^{\bar{\xi}  _{3} }\mathbf{F}^{H}\mathbf{\Lambda }^{\bar{\xi}  _{2} }\left ( \mathbf{F}^{-1}  \right ) ^{H}  \mathbf{\Lambda} ^{\bar{\xi}  _{1} } \\
        = \mathbf{I}.
	\end{array}
\end{equation}
Therefore, $O_{\rm{CM-CC-CM-GLCT} }^{\left ( a,b,c,d \right ) }$ is also a unitary matrix.

\subsection{Special Cases}
In this subsection, we discuss some special cases of CM-CC-CM-GLCT, including GFRFT and discrete Fresnel transform on graphs.

$\left ( 1 \right )$ Graph fractional Fourier transform 

When the parameter matrix is $\mathbf{M} = 
\begin{pmatrix}
	\cos\alpha  &\sin\alpha   \\
	-\sin \alpha &\cos\alpha  
\end{pmatrix}$, CM-CC-CM-GLCT degenerates into GFRFT with rotation angle parameter $\alpha$:
\begin{equation}
	\begin{array}{lcl}
		O_{\rm{GFrFT}}^{\alpha }\left ( \mathbf{x} \right )& = &O_{\rm{CM-CC-CM-GLCT} }^{\left ( \cos \alpha ,\sin \alpha ,-\sin \alpha ,\cos \alpha  \right ) }\left ( \mathbf{x} \right ) \\
		& = & \mathbf{\Lambda }^{\frac{\cos \alpha -1}{\sin \alpha } }\mathbf{F}^{-1}\mathbf{\Lambda} ^{-\sin \alpha }\mathbf{F\Lambda} ^{\frac{\cos \alpha -1}{\sin \alpha } }\mathbf{x}\\
		& = & \mathbf{\Lambda} ^{-\tan \frac{\alpha }{2} }\mathbf{F}^{-1}\mathbf{\Lambda} ^{-\sin \alpha }\mathbf{F\Lambda} ^{-\tan \frac{\alpha }{2}  }\mathbf{x}.
	\end{array}
\end{equation}
In \cite{zaktas1996DigitalCO}, the author proposed two digital calculation methods of FrFT. Among them, the first method is putting forward a concept similar to that in $\left ( 36 \right )$. So FrFT is decomposed into CM-CC-CM.

$\left (2 \right )$ Discrete Fresnel transform on graphs

When the parameter matrix is $\mathbf{M} = 
\begin{pmatrix}
	1 & \lambda z \\
	0 & 1
\end{pmatrix}$, CM-CC-CM-GLCT degenerates into discrete Fresnel transform on graphs, also known as CC. One-dimensional Fresnel transform is expressed as
\begin{equation}
	\begin{array}{lcl}
		O_{\rm{G Fresne}l}^{\lambda ,z}\left ( \mathbf{x} \right )& = & O_{\rm{CM-CC-CM-GLCT} }^{\left ( 1,\lambda z,0,1 \right )  }\left ( \mathbf{x} \right ) \\
		& = & \mathbf{F}^{-1}\mathbf{\Lambda} ^{-\lambda z}\mathbf{Fx},
	\end{array}
\end{equation}
where $\lambda$ is the wavelength and $z$ is the propagation distance.

$\left ( 3 \right )$ Other discrete operations on graphs

GFT, IGFT and CM are also special cases of CM-CC-CM-GLCT whose parameter matrices are $\left ( 0,1,-1,0 \right )$, $\left ( 0,-1,1,0 \right )$ and $\left ( 1,0,\xi ,1 \right )$ respectively. The discrete versions of these operations are simply $\mathbf{F}$, $\mathbf{F}^{-1}$ and $\mathbf{\Lambda} ^{\xi } $, without CM-CC-CM-GLCT.

\section{Comparison between CM-CC-CM-GLCT and CDDHFS-GLCT}
In this section, CM-CC-CM-GLCT is compared to the CDDHFs-GLCT in terms of computational complexity, additivity, and reversibility. Both be irrelevant to sampling periods and without oversampling operation.
\subsection{Computational complexity}
Recall the matrix form of the CDDHFs-GLCT showm in (15), that is, $O_{\rm{CDDHFs-GLCT} }^{\left ( a,b,c,d \right ) }=\mathbf{\Lambda} ^{\xi }\mathbf{Q}_{\delta }\mathbf{\Lambda} ^{\alpha }\mathbf{Q}^{\rm{T}}$. Each of the real matrix $\mathbf{Q}_{\delta }$ and $\mathbf{Q}^{\rm{T} }$ generates $2N^{2} $ real multiplications and each of complex diagonal matrix $\mathbf{\Lambda }^{\xi }$ and $\mathbf{\Lambda} ^{\alpha  }$ generates $N$ complex multiplications. Therefore, the computation totally includes $4N^{2}+8N $ real multiplications. Regarding CM-CC-CM-GLCT, firstly consider the $b\ne 0$ case. The matrix form of the CM-CC-CM-GLCT shown in (21), i.e., $O_{\rm{CM-CC-CM-GLCT} }^{\left ( a,b,c,d \right ) }=\mathbf{\Lambda} ^{\xi _{1} }\mathbf{F}^{-1}\mathbf{\Lambda} ^{\xi _{2} }\mathbf{F\Lambda} ^{\xi _{3} }$. There are three graph CMs which require $3N$ complex multiplications. GFT and IGFT can be calculated by FFT, which requires  $\left ( N/2 \right )\log_{2}{N}$ complex multiplications respectively. Therefore, the computational complexity of CM-CC-CM-GLCT for $b\ne 0$ is $12N+4N\log_{2}{N}$ real multiplications. CM-CC-CM-GLCT for $b = 0$ shown in (28) and (29), which contains one more GFT (or IGFT) than the $b \ne 0$ case. Therefore, its complexity increases to $12N+6N\log_{2}{N}$ real multiplications. Finally, we comprehensively compare the computational complexity in Table 1.

\begin{table}[]
	\caption{Complexity of CDDHFs-GLCT and CM-CC-CM-GLCT}
	\centering
	\resizebox{\textwidth}{!}{
	\begin{threeparttable}
	\begin{tabular}{|l|l|l|}
		\hline
		& Matrix form                                                                                                                                                                                                                                                                                      & Complexity                         \\ \hline
		CDDHFs-GLCT                            & $\mathbf{\Lambda}^{\xi}\mathbf{Q}_{\delta}\mathbf{\Lambda}^{\alpha}\mathbf{Q}^{\rm{T}}$                                                                                                                                                                                                                                                   & $4N^{2}+8N$                        \\ \hline
		CM-CC-CM-GLCT $(b\ne 0)$               & $\mathbf{\Lambda}^{\xi_{1}}\mathbf{F}^{-1}\mathbf{\Lambda}^{\xi_{2}}\mathbf{F\Lambda}^{\xi_{3}}$                                                                                                                                                                                                                                     & $12N+4N\log_{2}{N}$                  \\ \hline
		\multirow{2}{*}{CM-CC-CM-GLCT $(b=0)$} & \multirow{2}{*}{\begin{tabular}[c]{@{}l@{}}$\sqrt{-j}\mathbf{F\Lambda}^{\eta_{1}}\mathbf{F}^{-1}\mathbf{\Lambda}^{\eta_{2}}\mathbf{F\Lambda}^{\eta_{3}}$\\ $\sqrt{j}\mathbf{\Lambda}^{\mu_{1}}\mathbf{F}^{-1}\mathbf{\Lambda}^{\mu_{2}}\mathbf{F\Lambda}^{\mu_{3}}\mathbf{F}^{-1}$\end{tabular}} & \multirow{2}{*}{$12N+6N\log_{2}{N}$} \\
		&                                                                                                                                                                                                                                                                                                  &                                    \\ \hline
	\end{tabular}
	
	\begin{tablenotes}
		\footnotesize
		\item[*] The complexity of matrix  eigendecomposition for $\mathbf{Q}_{\delta }$ and $\mathbf{Q}^{\rm{T} }$ is not included.
	\end{tablenotes}
\end{threeparttable}
}
\end{table}

\subsection{Additivity property}
The additivity property of the GLCT is expressed as matrix multiplication, and the additivity property of the GLCT represents
\begin{equation}
	O_{\rm{GLCT}}^{\mathbf{M}_{1} }O_{\rm{GLCT}}^{\mathbf{M}_{2} }=O_{\rm{GLCT}}^{\mathbf{M}_{1}\times \mathbf{M}_{2}  },
\end{equation}
where $\mathbf{M}_{1}=\begin{pmatrix}
	a_{1} &b_{1}  \\
	c_{1}&d_{1}
\end{pmatrix}  $, $\mathbf{M}_{2}=\begin{pmatrix}
a_{2} &b_{2}  \\
c_{2}&d_{2}
\end{pmatrix}  $.

In the following, we compare the NMSE of the additivity property for CDDHFs-GLCT and CM-CC-CM-GLCT \cite{Pei2016FastDL}:
\begin{equation}
	{\rm NMSE}=\frac{\sum_{n=0}^{N-1}\left | O_{\rm{GLCT}}^{\mathbf{M}_{1}\times \mathbf{M}_{2} }\left ( x\left [ n \right ]  \right )  - O_{\rm{GLCT}}^{\mathbf{ M}_{1}}O_{\rm{GLCT}}^{\mathbf{M}_{2}}\left ( x\left [ n \right ]  \right )    \right |^{2} }{\sum_{n=0}^{N-1}\left | O_{\rm{ GLCT}}^{\mathbf{M}_{1}\times \mathbf{M}_{2} }\left ( x\left [ n \right ]  \right )   \right |^{2}   } .
\end{equation}
The simulation is carried out on graph signals. Eight kinds of graph signals are shown in Figure 1 and described as follows: \\
(1) $\mathbf{x}_{1}$: The bipolar rectangular signal on the random regualar graph. The random regualar graph is generated by randomly generating 260 positions and by 3- nearest neighbor process. \\
(2) $\mathbf{x}_{2}$: The bipolar rectangular signal on the sprial graph. The sprial graph consists of 160 nodes and 930 vertices. \\
(3) $\mathbf{x}_{3}$: The bipolar rectangular signal on the community graph. It is a 2 dimentional random sensor graph, which consists of 440 nodes and 8774 vertices. All the coordonates are between 0 and 1. \\
(4) $\mathbf{x}_{4}$: The bipolar rectangular signal on the sphere graph. The spherical-shaped graph creates a graph from points sampled on a hyper-sphere, which consists of 280 nodes and 3182 vertices. \\
(5) $\mathbf{x}_{5}$: The bipolar rectangular signal on the sensor graph. It is a 2 dimentional random sensor graph, which consists of 260 nodes and 1854 vertices. All the coordonates are between 0 and 1. \\
(6) $\mathbf{x}_{6}$: The bipolar rectangular signal on the swiss roll graph. The swiss roll graph consists of 200 nodes and 1444 vertices.\\
(7) $\mathbf{x}_{7}$: The bipolar rectangular signal on the commet graph. The commet graph is a simple path graph with a star degree 30 at its end, which consists of 60 nodes and 118 vertices. \\
(8) $\mathbf{x}_{8}$: The bipolar rectangular signal on the path graph. The path graph consists of 50 nodes and 98 vertices. \\
The parameters in $\mathbf{M}_{1}$ and $\mathbf{M}_{2}$ are random numbers uniformly distributed on the interval $\left [ -2,2 \right ]$, which are used for 50 simulation runs. The additivity NMSE results in (39) is obtained. The NMSEs sorted in ascending order using $\mathbf{x}_{1}-\mathbf{x}_{8}$ as input signals are shown in Figure 2 respectively. In addition, the numerical verification of the average value of NMSEs for these graph signals for running 1000 times is shown in Table 2. These examples reveal that CM-CC-CM-GLCT has similar performance to the CDDHFs-GLCT in the additivity property.
\begin{figure}
\subfigure[]{
	\includegraphics[scale=0.35]{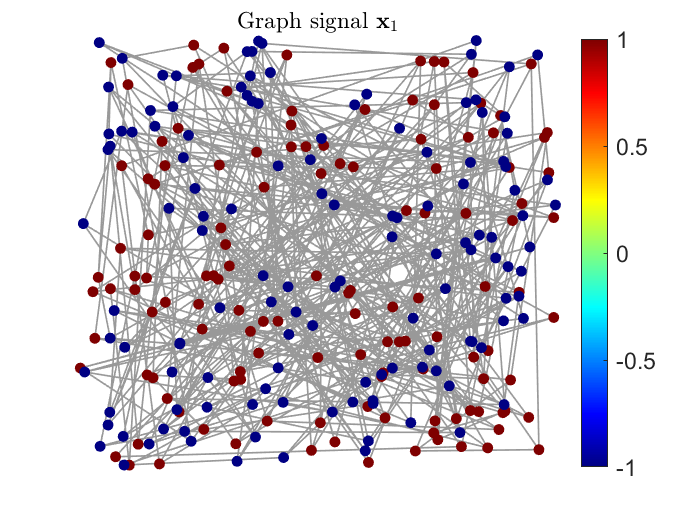} \label{1}
}
\quad
\subfigure[]{
	\includegraphics[scale=0.35]{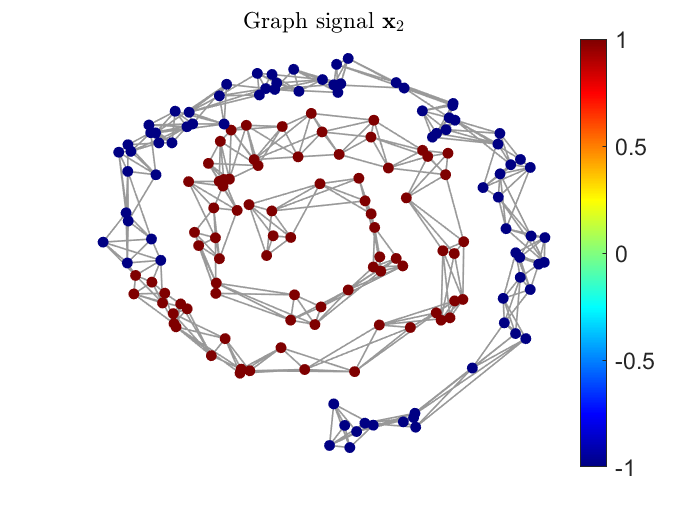} \label{2} 
}
\quad
\subfigure[]{
	\includegraphics[scale=0.35]{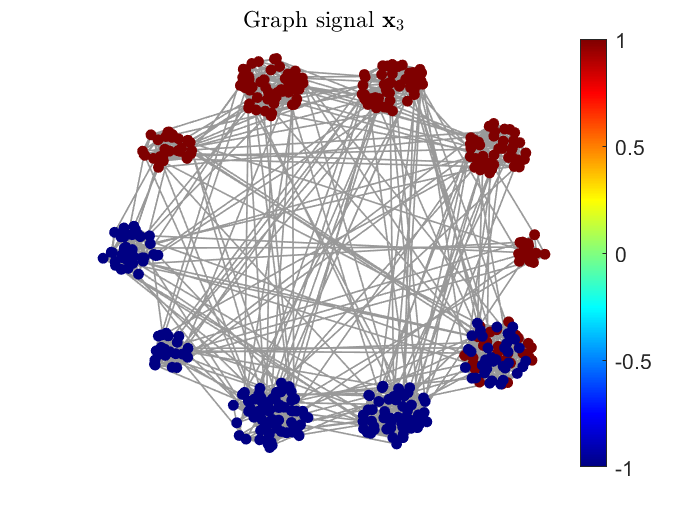} \label{3} 
}
\quad
\subfigure[]{
	\includegraphics[scale=0.35]{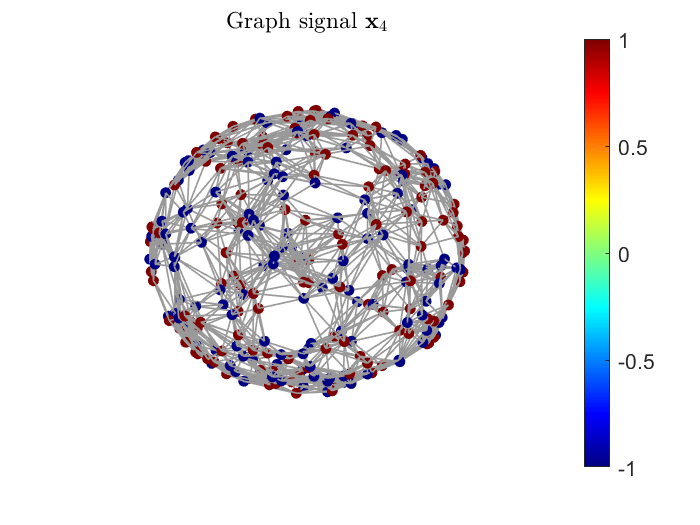} \label{4} 
}
\quad
\subfigure[]{
	\includegraphics[scale=0.35]{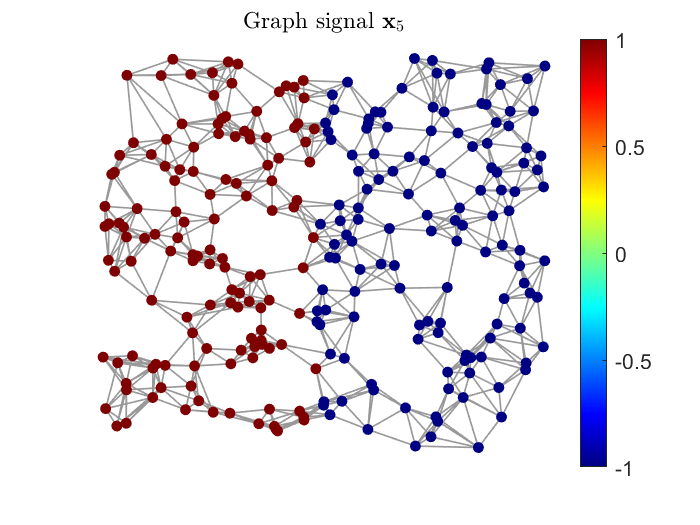} \label{5}
}
\quad
\subfigure[]{
	\includegraphics[scale=0.35]{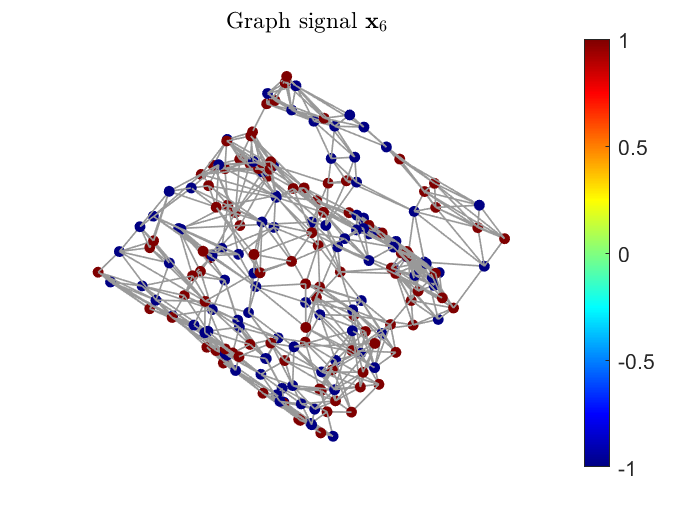} \label{6} 
}
\quad
\end{figure}
\begin{figure}
	\subfigure[]{
		\includegraphics[scale=0.35]{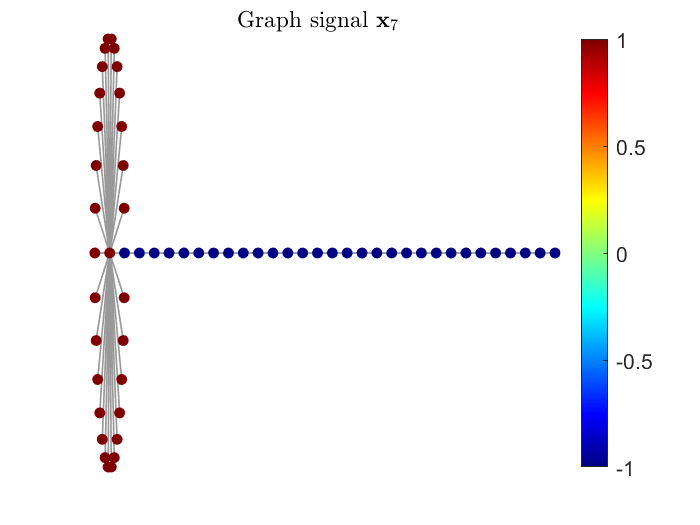} \label{1}
	}
	\quad
	\subfigure[]{
		\includegraphics[scale=0.35]{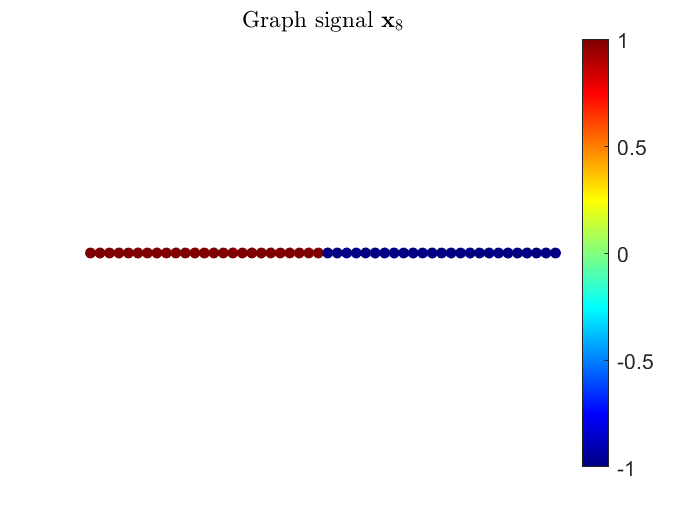} \label{1}
	}
	\quad
	\caption{Graph signals $\mathbf{x}_{1}$ to $\mathbf{x}_{8}$.}
\end{figure}
\begin{figure}
\subfigure[]{
	\includegraphics[scale=0.35]{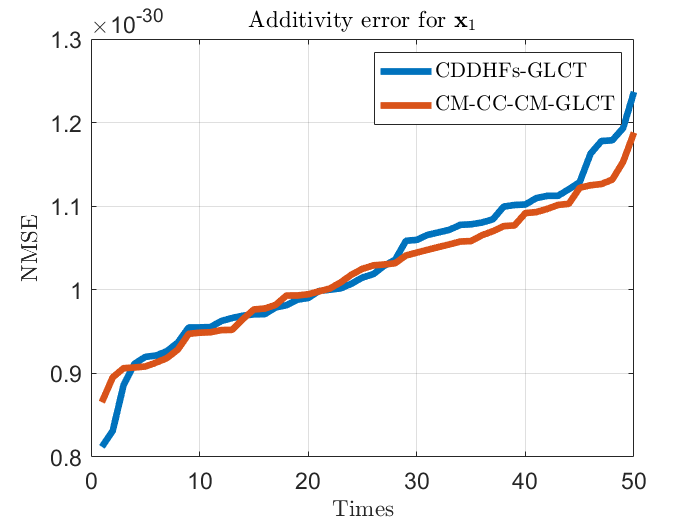} \label{1}
}
\quad
\subfigure[]{
	\includegraphics[scale=0.35]{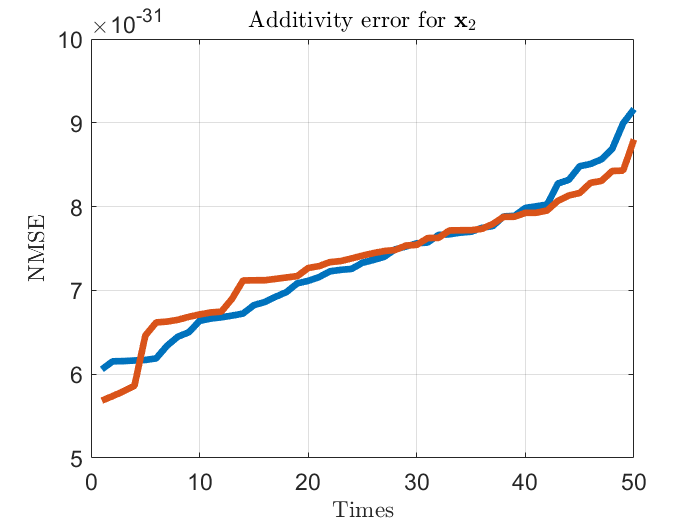} \label{2}
}
\quad
\subfigure[]{
	\includegraphics[scale=0.35]{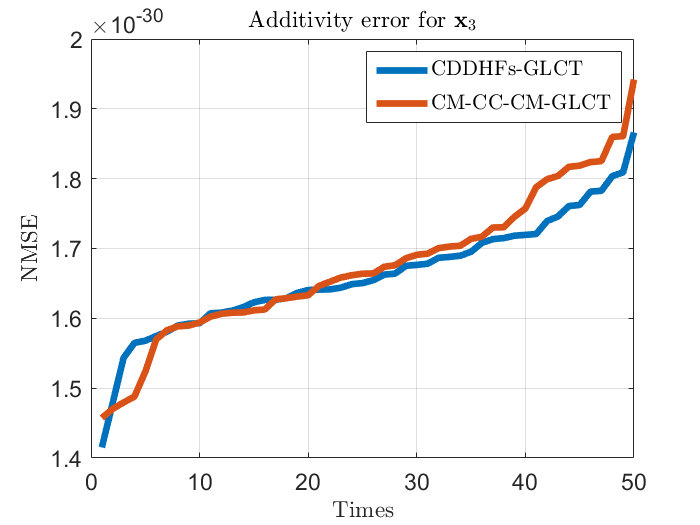} \label{3}
}
\quad
\subfigure[]{
	\includegraphics[scale=0.35]{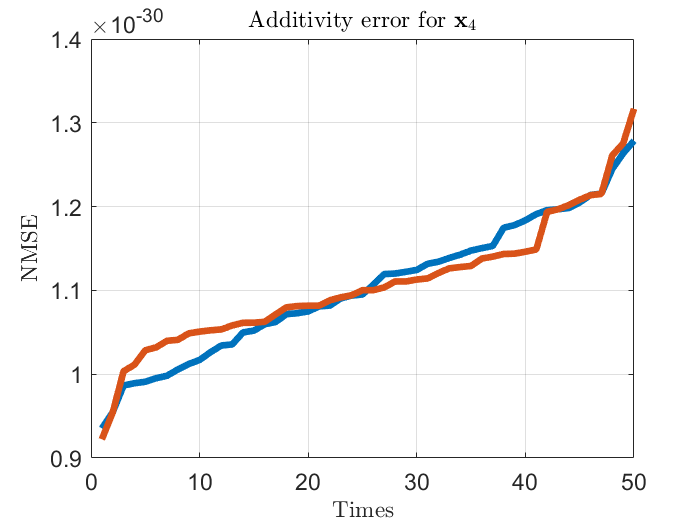} \label{4}
}
\quad
\subfigure[]{
	\includegraphics[scale=0.35]{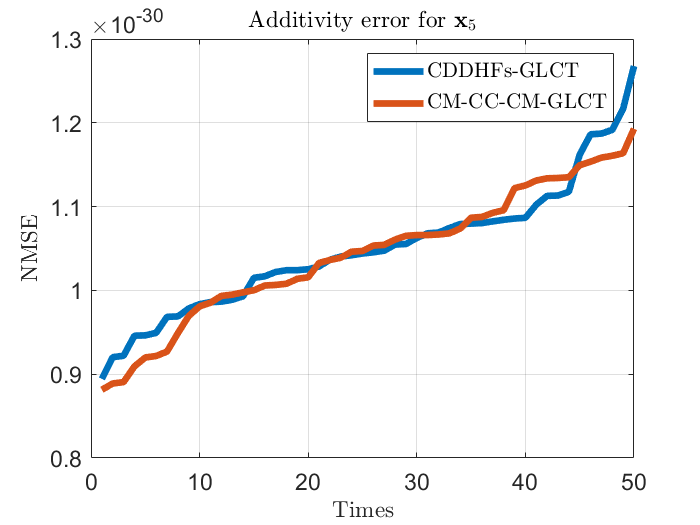} \label{5}
}
\quad
\subfigure[]{
	\includegraphics[scale=0.35]{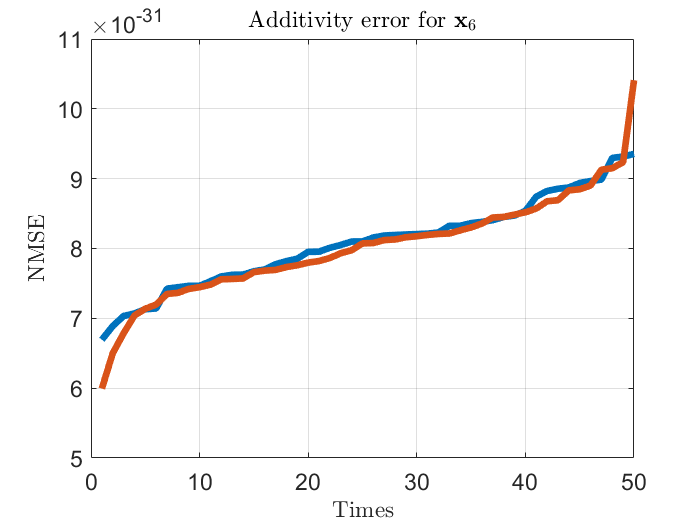} \label{6}
}
\quad
\end{figure} 
\begin{figure}
	\subfigure[]{
		\includegraphics[scale=0.35]{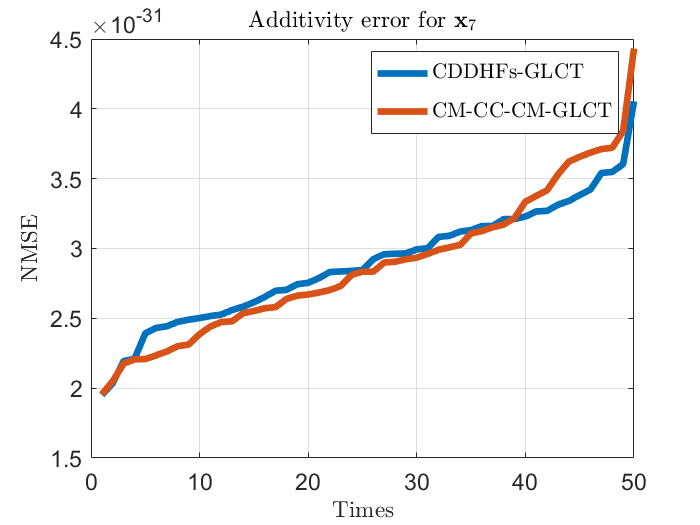} \label{1}
	}
	\quad
		\subfigure[]{
		\includegraphics[scale=0.35]{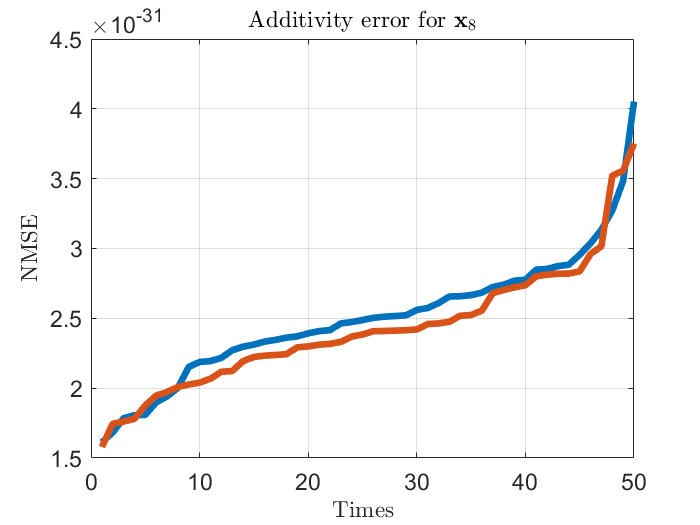} \label{1}
	}
	\quad
	\caption{Normalized mean-square errors (NMSEs) of the additivity property for 50 different sets of $\mathbf{M}_{1}$ and $\mathbf{M}_{2}$.}
\end{figure} 

\subsection{Revertibility property}
Next, we compare the reversibility of CM-CC-CM-GLCT and the CDDHFs-GLCT, by checking the NMSE of reversibility \cite{Pei2016FastDL}:
\begin{equation}
	{\rm NMSE}=\frac{\sum_{n=0}^{N-1}\left | x\left [ n \right ] -O_{\rm{GLCT}}^{\mathbf{M}^{-1} }O_{\rm{GLCT}}^{\mathbf{M}} \left ( x\left [ n \right ]  \right )    \right |^{2}   }{\sum_{n=0}^{N-1}\left | x\left [ n \right ]   \right |^{2}   }.
\end{equation}

The parameters in $\mathbf{M}$ are random numbers uniformly distributed on the interval [-2, 2], which are used for 50 simulation runs. The reversibility NMSE results in (40) is obtained. The NMSEs sorted in ascending order using $\mathbf{x}_{1}-\mathbf{x}_{8}$ as input signals are shown in Figure 3 respectively. As shown in figures, all NMSEs of CM-CC-CM-GLCT are below. In addition, the numerical verification of the average value of NMSEs for these graph signals for running 1000 times is shown in Table 3. If GLCT is reversible, there is no necessity to develop the inverse GLCT since it can be realized by the forward GLCT. Finally, Table 4 summarizes the comparison between the CDDHFs-GLCT and CM-CC-CM-GLCT.
\begin{table}[]
	\caption{Comparison of 1000 additivity running averages}
	\centering
	\begin{threeparttable}
		\resizebox{\textwidth}{!}{
	\begin{tabular}{|l|l|l|l|l|l|l|l|l|}
		\hline
		& $\mathbf{x}_{1}$ & $\mathbf{x}_{2}$ & $\mathbf{x}_{3}$ & $\mathbf{x}_{4}$ & $\mathbf{x}_{5}$ & $\mathbf{x}_{6}$ & $\mathbf{x}_{7}$ & $\mathbf{x}_{8}$ \\ \hline
		CDDHFs-GLCT   & 8.7829           & 16.8790          & 7.4394           & 11.1150          & 8.9161           & 8.1654           & 2.9040           & 2.4280           \\ \hline
		CM-CC-CM-GLCT & 8.7466           & 16.6340          & 7.4007           & 11.1390          & 8.8399           & 8.0253           & 2.8740           & 2.3802           \\ \hline
	\end{tabular}
		}
	\begin{tablenotes}
	\footnotesize
	\item[*] The order of magnitude is $e^{-31}$ in the table 2.
	\end{tablenotes}
\end{threeparttable}
\end{table}

\begin{figure}
\subfigure[]{
	\includegraphics[scale=0.35]{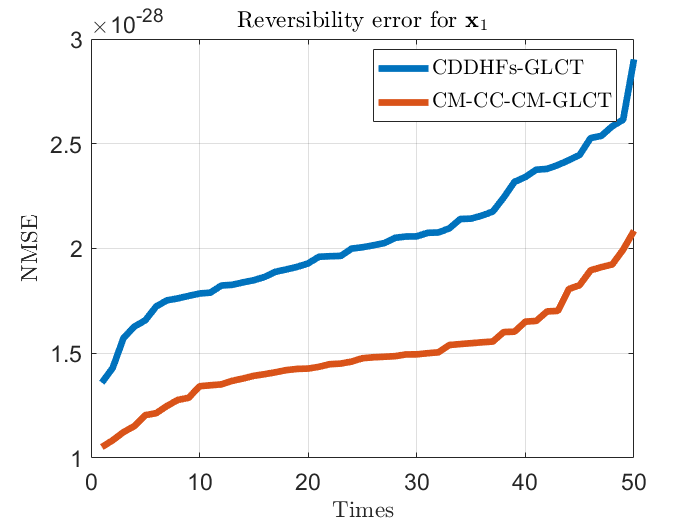} \label{1}
}
\quad
\subfigure[]{
	\includegraphics[scale=0.35]{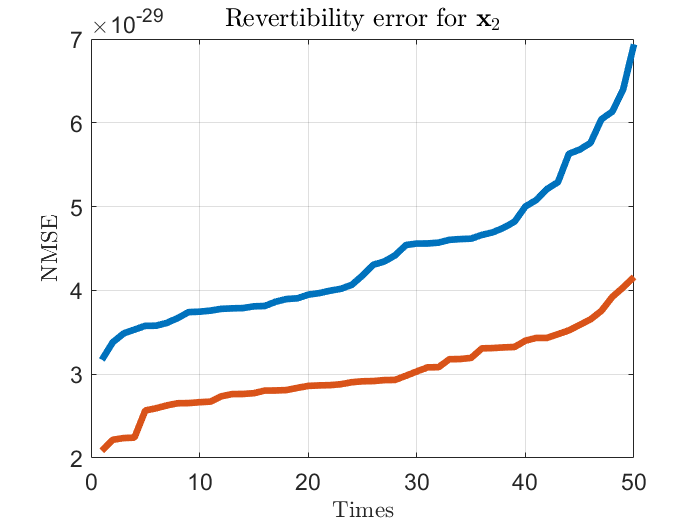} \label{2}
}
\quad
\subfigure[]{
	\includegraphics[scale=0.35]{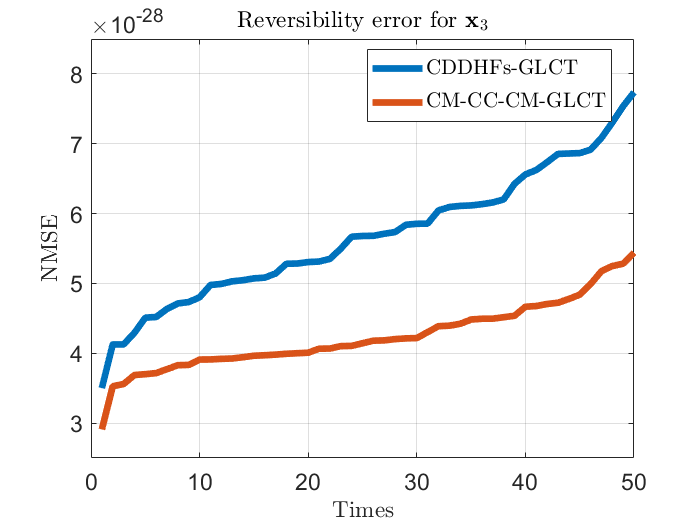} \label{3}
}
\quad
\subfigure[]{
	\includegraphics[scale=0.35]{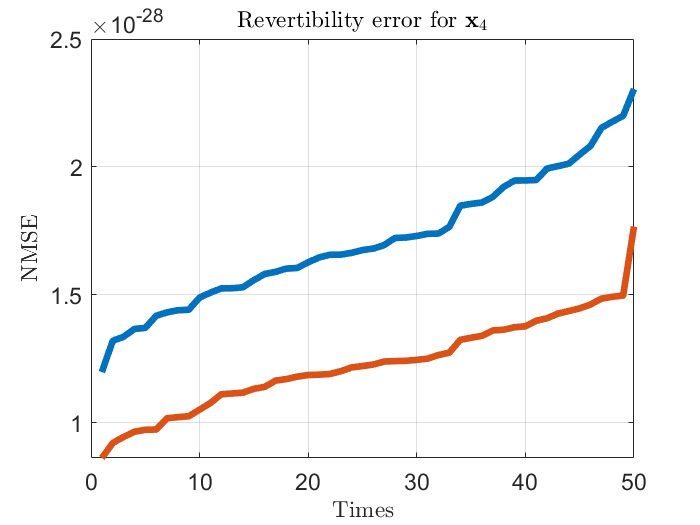} \label{4}
}
\quad
\subfigure[]{
	\includegraphics[scale=0.35]{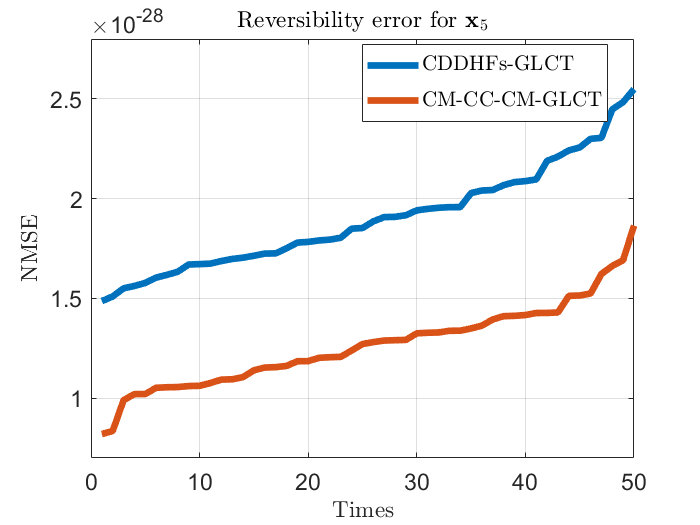} \label{1}
}
\quad
\subfigure[]{
	\includegraphics[scale=0.35]{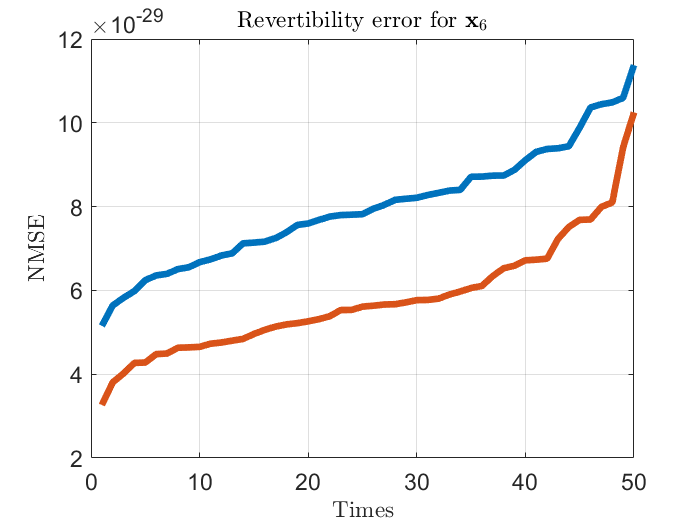} \label{2}
}
\quad
\end{figure} 
\begin{figure}[]
	\subfigure[]{
		\includegraphics[scale=0.35]{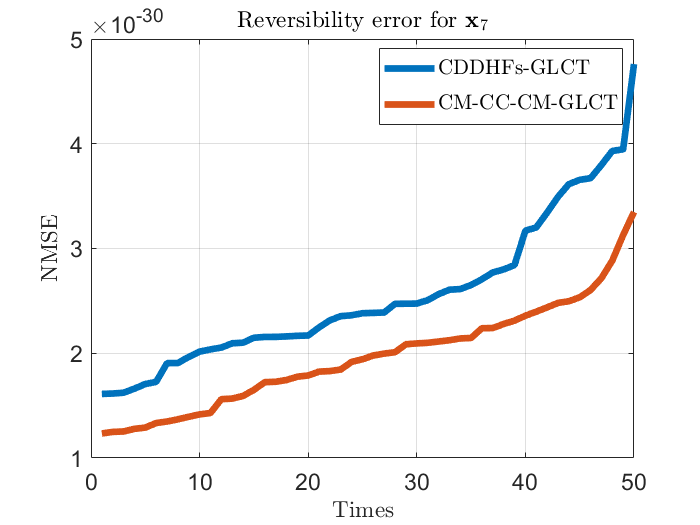} \label{1}
	}
	\quad
	\subfigure[]{
		\includegraphics[scale=0.35]{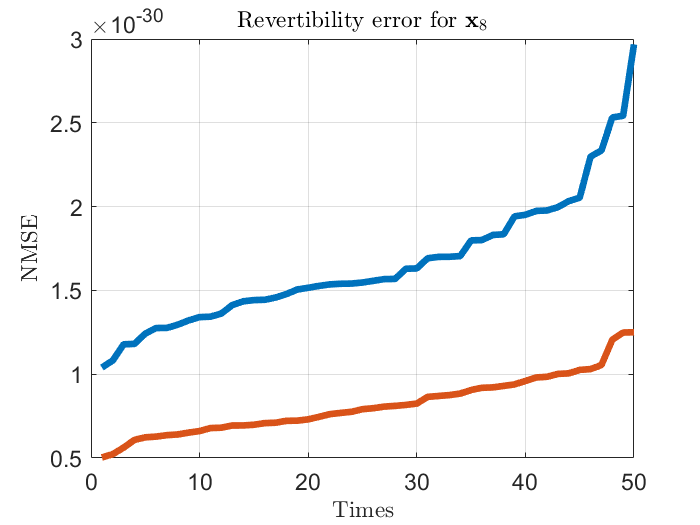} \label{1}
	}
	\quad
	\caption{Normalized mean-square errors (NMSEs) of the reversibility property for 50 different sets of $\mathbf{M}$.}
\end{figure}
\begin{table}[]
\caption{Comparison of 1000 reversibility running averages}
\centering
\begin{threeparttable}
	\resizebox{\textwidth}{!}{
		\begin{tabular}{|l|l|l|l|l|l|l|l|l|}
			\hline
			& $\mathbf{x}_{1}$ & $\mathbf{x}_{2}$ & $\mathbf{x}_{3}$ & $\mathbf{x}_{4}$ & $\mathbf{x}_{5}$ & $\mathbf{x}_{6}$ & $\mathbf{x}_{7}$ & $\mathbf{x}_{8}$ \\ \hline
			CDDHFs-GLCT   & 53.0580          & 54.4980          & 5.3073           & 20.0330          & 10.3350          & 8.1785           & 0.2506           & 0.1704           \\ \hline
			CM-CC-CM-GLCT & \textbf{50.0590} & \textbf{49.8210} & \textbf{4.7535}  & \textbf{18.1650} & \textbf{9.5250}  & \textbf{5.7364}  & \textbf{0.1995}  & \textbf{0.1264}  \\ \hline
		\end{tabular}
	}
	\begin{tablenotes}
		\footnotesize
		\item[*] The order of magnitude is $e^{-29}$ in the table 3.
	\end{tablenotes}
\end{threeparttable}
\end{table}
\begin{table}[]
\caption{Comparison between the CDDHFs-GLCT and CM-CC-CM-GLCT}
\centering
	\begin{tabular}{|l|l|l|}
		\hline
		& CDDHFs-GLCT                                                             & 
		CM-CC-CM-GLCT                                                      \\ \hline
		Complexity    & $O\left ( N^{2}  \right )$ & $O\left ( N\log_{2}{N}   \right )$ \\ \hline
		Additivity    & Approximate                                                               & Approximate                                                    \\ \hline
		Reversibility & Worse                                                                     & Better                                                         \\ \hline
	\end{tabular}
\end{table}

\section{Conclusions}
Graph signal processing extends conventional discrete signal processing to graph signals with topological structures, offering an effective approach for handling data with intricate structures. This technology finds widespread application. The GLCT, like the LCT, represents a versatile multi-parameter linear integral transformation with enhanced flexibility, making it a significant tool in signal processing. Based on the DLCT decomposition method and CM-CC-CM decomposition, this paper develops CM-CC-CM-GLCT which irrelevant to sampling periods and without oversampling operation. Various properties and special cases of the CM-CC-CM-GLCT are derived and discussed. In terms of computational complexity, additivity, and reversibility, CM-CC-CM-GLCT is compared to the previous CDDHFs-GLCT. Theoretical analysis demonstrates that the computational complexity of the CM-CC-CM-GLCT is obviously decreased. In addition, the normalized mean-square error is used for simulation. Simulation results indicate that CM-CC-CM-GLCT achieves similar additivity to the CDDHFs-GLCT. Notably, the CM-CC-CM-GLCT exhibits better reversibility.



\bibliographystyle{elsarticle-num}  
\bibliography{ref}  
\end{document}